\documentclass{iau} 
\usepackage{graphicx, natbib}

\title[] 
{High-Temperature Chemistry\\ in External Galaxies}

\author[N. Harada]   
{Nanase Harada$^1$
}

\affiliation{$^1$Academia Sinica Institute of Astronomy and Astrophysics 
\\ Taipei, Taiwan
\\ email: {\tt harada@asiaa.sinica.edu.tw}}

\pubyear{2017}
\volume{332}  
\jname{Astrochemistry VII -- Through the Cosmos from Galaxies to Planets}
\editors{Maria Cunningham, Tom Millar \& Yuri Aikawa eds.}

\begin{document}

\maketitle

\newcommand{\apj}{\textit{ApJ}}
\newcommand{\apjl}{\textit{ApJL}}
\newcommand{\aap}{\textit{A\&A}}
\newcommand{\aapl}{\textit{A\&AL}}
\newcommand{\pasj}{\textit{PASJ}}
\newcommand{\mnras}{\textit{MNRAS}}

\begin{abstract}
In external galaxies, some galaxies have higher activities of star formation and central supermassive black holes. 
The interstellar medium in those galaxies can be heated by different mechanisms such as UV-heating, X-ray heating, cosmic-ray heating, and shock/mechanical heating. 
Chemical compositions can also be affected by those heating mechanisms.
Observations of many molecular species in those nearby galaxies are now possible with the high sensitivity of Atacama Large Millimeter/sub-millimeter Array (ALMA).
Here I cover different chemical models for those heating mechanisms. 
In addition, I present recent ALMA results of extragalactic astrochemistry including our results of a face-on galaxy M83 and an infrared-luminous merger NGC 3256.
\end{abstract}
\firstsection 
\section{Astrochemistry in External Galaxies}

Some external galaxies are going through violent activities that Milky Way currently does not experience.
Star formation rates can vary orders of magnitude among galaxies. 
Active galactic nuclei (AGNs) emit high intensity of high-energy photons such as X-rays.
Merger events are thought to be important processes in the course of galaxy evolution.
Such phenomena can affect the properties of molecular clouds in those galaxies.
Elevated gas temperatures are one of those properties. 
Central regions of starburst galaxies and active galaxies tend to have higher gas kinetic temperature than Milky Way sources (Table \ref{tab:temp}).
There are various heating mechanisms of the molecular medium.
Starburst galaxies may have higher UV radiation field, while AGNs emit high luminosity of X-rays.
After the starburst activity, supernova explosions can also create high flux of cosmic-rays.
Both starburst and AGNs can cause high degree of turbulence or shocks.
Those heating sources affect the chemical composition in different ways
although some effects are similar.

Observing molecular tracers of those heating mechanisms has recently become easier.
After the commission of Atacama Large Millimeter/sub-millimeter Array (ALMA) in 2011, 
its early science has already shown its capability to widely expand the potential 
of spatially-resolved astrochemical studies in external galaxies.
 ALMA has increased the number of detectable species,
 and its arcsecond/sub-arcsecond resolution 
 allows comparisons of various locations in galaxies in a scale of giant molecular clouds (GMCs).
 A larger number of galaxies can be studied in astrochemical point of view with ALMA,
 and comparisons between various galaxy types become possible.
 Since some galactic nuclei are highly obscured, it would be useful for a wider astronomical community if some molecular
 species are found to be tracers of star formation or AGN activities.

In this proceedings, I will summarize previous theoretical models for each heating mechanisms
and observational results especially ALMA results in galaxies affected by those heating sources.

\begin{table}
  \begin{center}
  \caption{Gas kinetic temperature in selected galaxies or Galactic sources traced by excitations of ammonia (NH$_{3}$) or formaldehyde (H$_2$CO). 
  IRDC stands for an infrared dark cloud.
  Two temperatures are listed if multi-components are derived from their observations.
  Beam sizes are listed because hot components tend to be compact, and higher angular resolution observations are more likely to pick up 
  hot components if there are any. References are a: \citet{2011ApJ...736..163R}, b:  \citet{2016A&A...586A..50G}, c: \citet{2013ApJ...779...33M},
  d: \citet{2011A&A...529A.154A}}
  \label{tab:temp}
 {
  \begin{tabular}{cccccc}\hline 
Source & Temperature (K) & Galaxy type & Molecule & Beam size (pc) & Reference\\
\hline
Milky Way IRDC &8-13&---&NH$_3$&0.08 - 0.16&a\\
Milky May CMZ &60 / $>100$ &---&H$_2$CO&1.2&b \\
NGC 253 &78 $\pm$ 22 / $>100$&Starburst&NH$_3$&500&c\\
Arp 220 &234 $\pm$ 52&Starburst (+ AGN?)&NH$_3$&12000&c\\
NGC 1068 &$80\pm20$/$140\pm30$&AGN&NH$_3$&1200&d\\
\hline
  \end{tabular}
  }
 \end{center}
\vspace{1mm}
\end{table}

\section{Heating Mechanisms and Theoretical Models}


\subsection{Photon-dominated regions}
Starburst galaxies may have orders of magnitude higher star formation rate than the Milky Way.
In those galaxies, the interstellar radiation field could also increase by orders of magnitude.
In the vicinity of massive stars, high-energy UV-photons create ionized HII regions. 
Further out, after photons with energies above the ionization energy of hydrogen (13.6 eV) are absorbed,
there are photon-dominated regions (PDRs) where atoms with lower ionization energy than that of hydrogen (e.g., carbon) can be ionized.
The interstellar medium (ISM) in PDRs is heated by photo-electric effect,
de-excitation of vibrationally excited species after the photopumping of H$_{2}$,
and IR pumping of the OI line and its collisional de-excitation \citep{2005pcim.book.....T}. 
Those UV-photons not only ionize, but dissociate molecules.
Therefore, more ions or radicals are found in PDRs.
The rate coefficients of photo-dissociation or photo-ionization scales as $k \propto exp(-bA_{\rm V})$ with typical values of $b = 1-4$,
the effects of UV-photons are not significant when the molecular clouds are shielded by columns of $N_{\rm H} \gtrsim 2 \times 10^{22}$ cm$^{-2}$.
Many PDR models using a one-dimensional slab have been developed over decades \citep[][and references therein]{2007A&A...467..187R}
to include complex physics to radiative transfer and thermal balance. 
In extragalactic observations of molecular lines where the spatial scale is 
at least GMC-scale, how much emission comes from PDRs needs to be investigated 
with the consideration of the density probably distribution of the ISM
(see also Section \ref{sec:gmc}).
PDR models including 3-dimensional structure of the molecular clouds \citep[e.g., ][]{2012A&A...544A..22L}
are useful, and such models in the context of starburst galaxies are needed.

\subsection{X-ray dominated regions}
X-rays from AGNs are another possible heating source if the galactic nuclei are active.
They can heat the medium by Coulomb heating, or ionization or excitation 
that eventually lead to heating the gas \citep{m96}.
X-ray heating is more efficient than UV-photon heating given the same amount of energy \citep{2005A&A...436..397M} .
Beside heating and ionization, it can dissociate molecules via internally generated UV-photons.
 X-rays can also produce doubly-ionized ions such as C$^{++}$ and S$^{++}$, and may increase the fractional abundances of species 
 produced from those species \citep{2008ApJ...675L..81A}.
Since the photo-absorption cross sections is $1.0 \times 10^{-24}$ cm$^{-2}$ for 10 keV photon with the energy 
dependence of $E^{- \gamma}$ where $\gamma = 2.6 - 3$,
 X-rays penetrate up to $\sim 10^{24} - 10^{25}$ cm$^{-2}$ \citep{m96}.
This means that in highly-obscured galactic nuclei, the radius of influence of X-rays
will be dependent on the morphology of the surrounding ISM. 
\citet{2005A&A...436..397M} and \citet{2007A&A...461..793M} has compared the effects of PDR and XDR using a steady-state model
in a one-dimensional slab model to derive fractional abundances and intensities of molecular lines.

\begin{figure*}[bh]
 \centering{
\includegraphics[width=0.35\textwidth, trim= 0 0 0 0, angle=90]{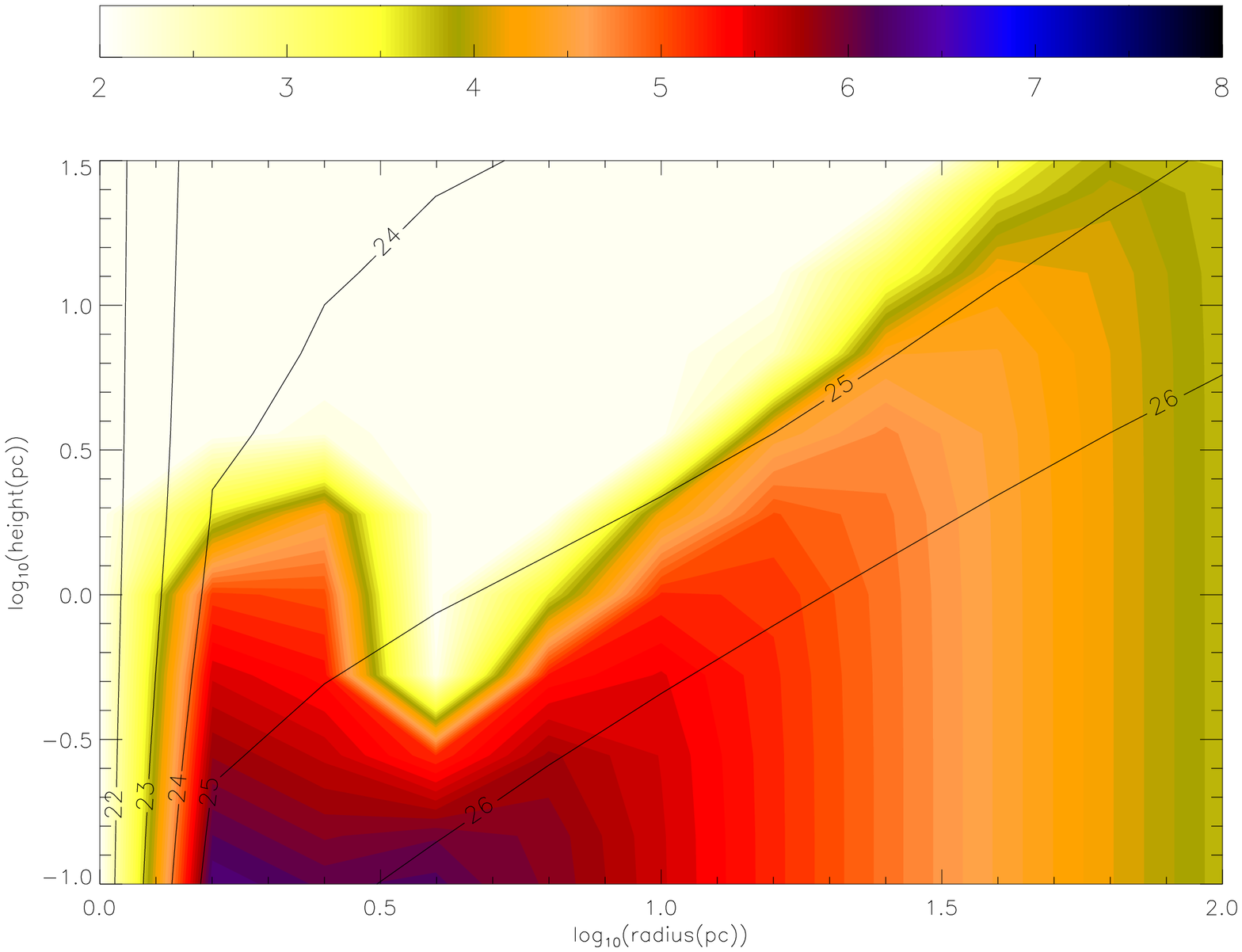}
\includegraphics[width=0.35\textwidth, trim= 0 0 0 0,angle= 90]{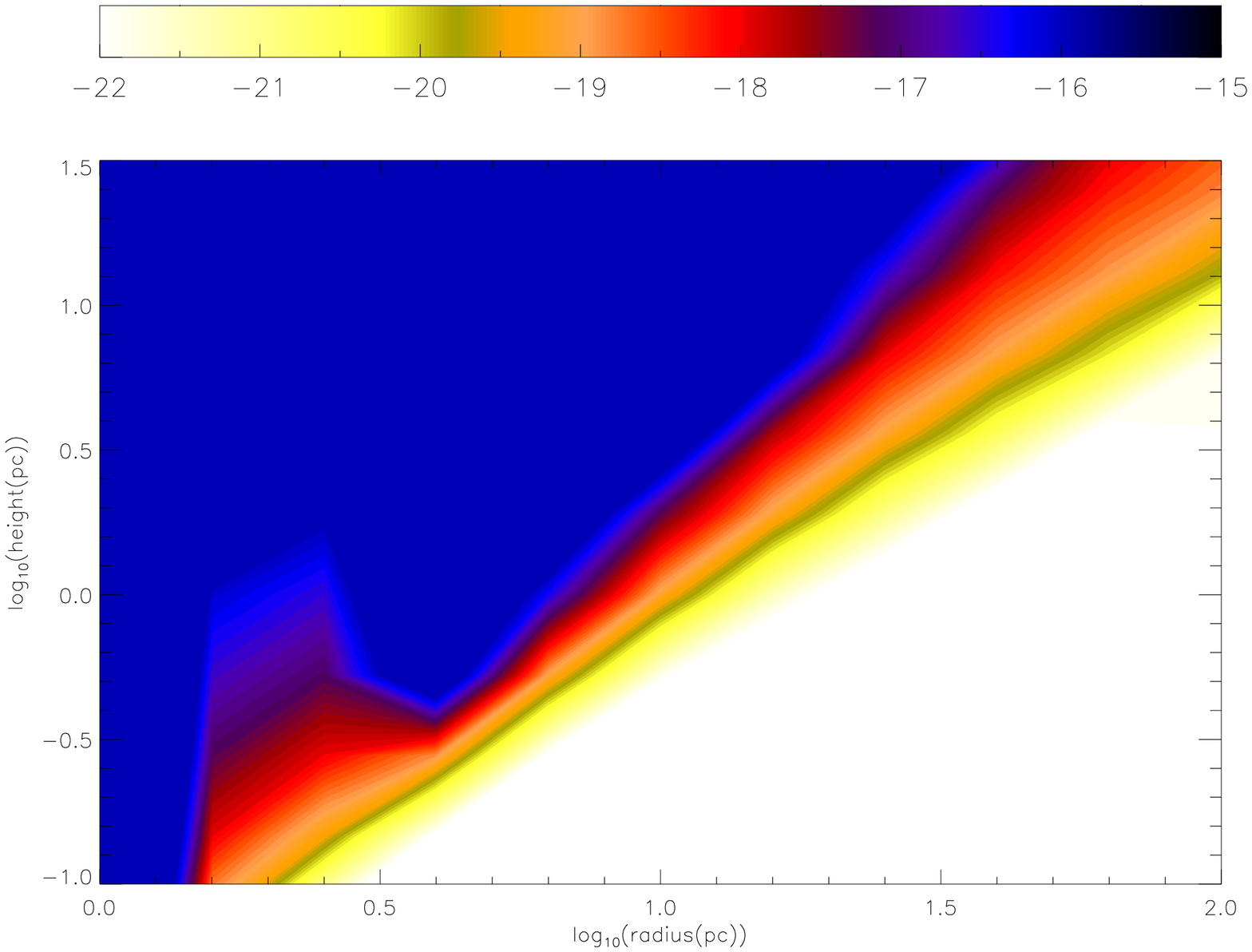}}
\centering{
\includegraphics[width=0.35\textwidth, trim= 0 0 0 0, angle=90]{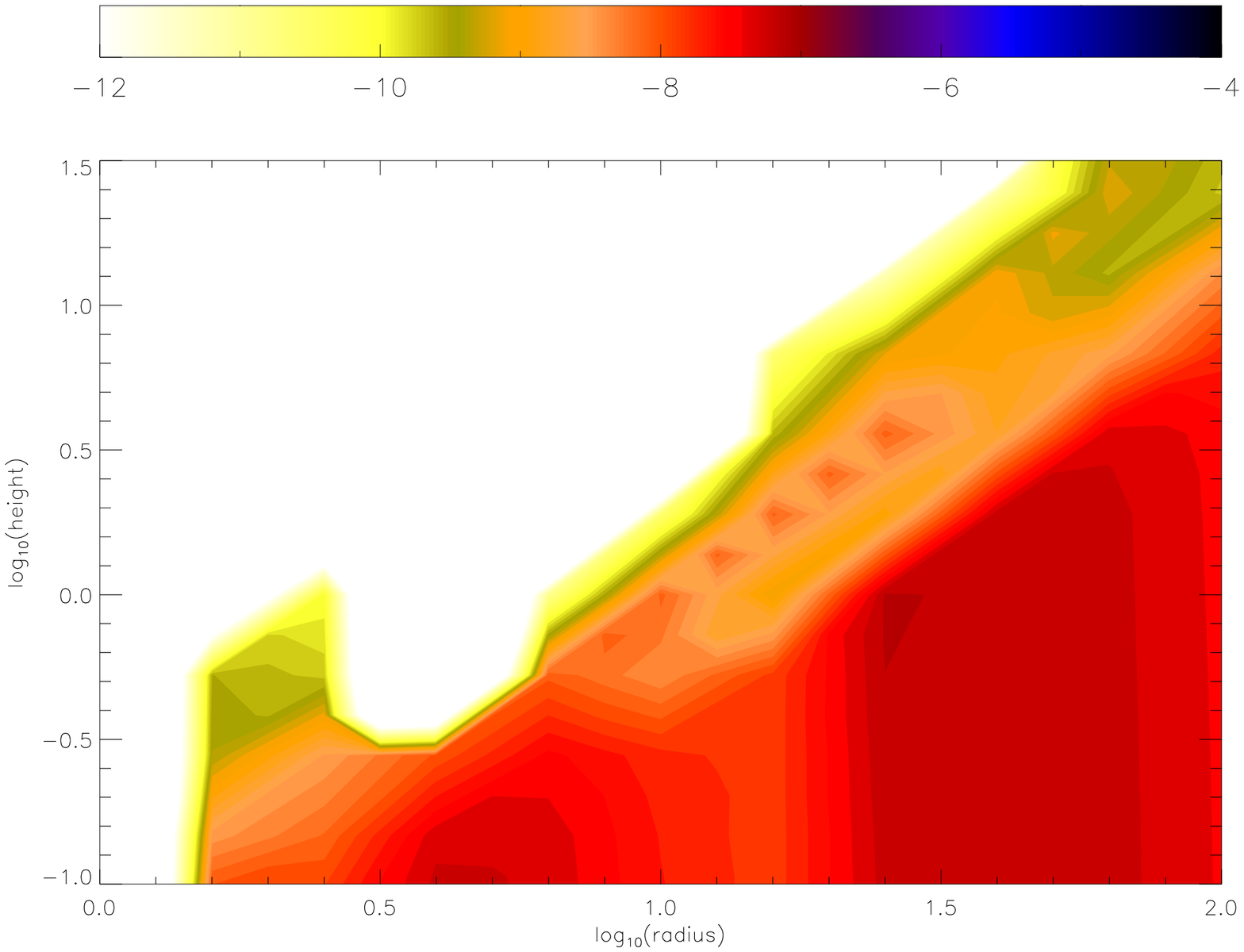}
\includegraphics[width=0.35\textwidth, trim= 0 0 0 0,angle= 90]{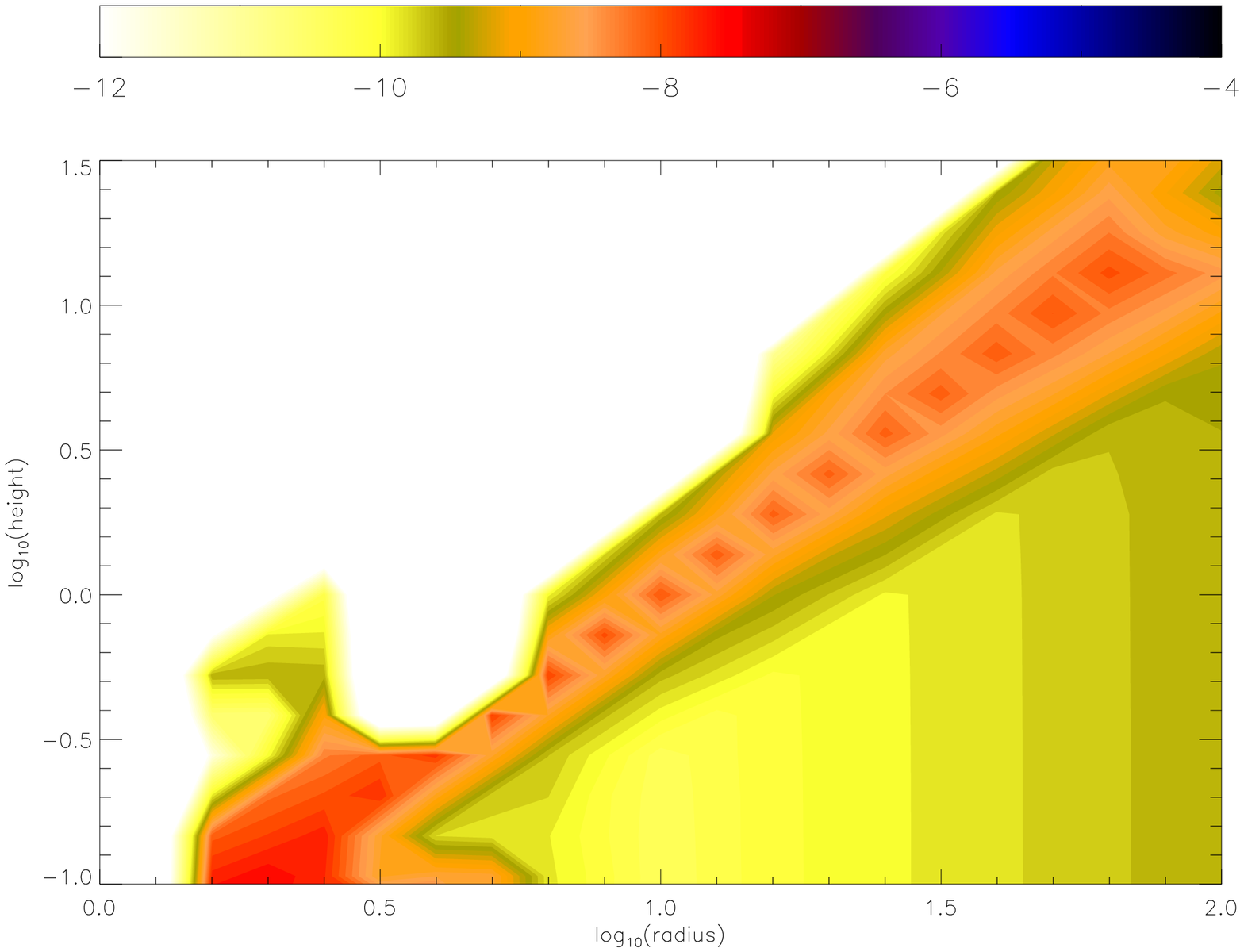}}
\caption{($Top~left$) The density structure of the AGN disk model used in \citet{2013ApJ...765..108H}. 
Column densities intervening X-rays from the central AGN are also shown as black contours. 
($top~right$) Values of $\zeta_{X} /n$ where $\zeta_{X}$ is the X-ray ionization rate and 
$n$ is the density. The influence of X-rays should roughly scale as $\zeta_{X} /n$ \citep{1996A&A...306L..21L}.
Fractional abundances of HCN in an axisymmetric model around an AGN in \citet{2013ApJ...765..108H} at
($bottom~left$) a dynamical time and ($bottom~right$) steady-state. }\label{fig:hcn_agn}
\end{figure*}

Although one-dimensional models are useful in grasping the dependence of the chemistry on the X-ray strength,
how far the effects of X-rays can extend can be modeled should be examined though a multi-dimensional model.
Time dependance is also important for some species.
\citet{2013ApJ...765..108H} conducted a chemical model with a time-dependent gas-phase chemical model
in axisymmetric disk models implied and modified from the density structure in \citet{2005ApJ...630..167T}. 
In these models, the importance of time-dependence was shown for species such as HCN and CN (Figure \ref{fig:hcn_agn}).
Although the steady-state is achieved in a relatively short time scale for high X-ray flux regions,
 the accurate comparisons between the regions with and without strong influence of X-rays should be made with the consideration of time dependence.
These models also showed the variation of the distribution of XDRs in different density structure of models.

\subsection{Cosmic-ray-dominated regions}
Cosmic rays from supernova explosions cause ionization further into molecular clouds than the UV-photons or X-rays,
up to a column of $N_{\rm H} \sim 10^{26}$ cm$^{-2}$ \citep{1981PASJ...33..617U}.
Similar to X-rays, cosmic rays can also internally generate UV-photons which can dissociated molecules.
A typical value of a cosmic-ray ionization rate in dense cores in the spiral arm part of the Galaxy ranges $(1 - 5) \times 10^{-17}\,$s$^{-1}$.
 This value is found to be higher in Galactic diffuse clouds because low-energy cosmic rays have higher cross sections,
  causing about one order of magnitude higher ionization rate in diffuse clouds while they become attenuated before reaching into the dense clouds.
  \citep{2009ApJ...694..257I,2009A&A...501..619P}.
In starburst galaxies, star formation rates per unit volume are higher than the Galaxy,
and cosmic-ray ionization rates are also expected to increase due to the higher supernova rates. 
According to \citet{2010ApJ...720..226P}, for the case of extreme starburst galaxies such as ultra-luminous infrared galaxies (ULIRGs; $L_{\rm IR} > 10^{12} L_{\odot}$), 
cosmic-ray energy density can be 3-4 orders of magnitude higher than the Galaxy, and the cosmic-ray
heating alone can raise the gas kinetic temperature to 80 - 160 K. 
\citet{2006ApJ...650L.103M} compared the scenarios between an XDR model and a CRDR+PDR model,
to find the elevated CI emission in an XDR model.
\citet{2011MNRAS.414.1583B} also modeled the chemistry in the CRDR+PDR model, 
but for larger number of molecular species. Their results showed that OH, H$_2$O, H$_3^+$ , H$_3$O$^+$ and OH$^+$
have high fractional abundances in high cosmic-ray ionization rate ($\zeta > 10^{-13}$ s$^{-1}$).
\citet{2011A&A...525A.119M} modeled the CRDR with the consideration of mechanical-heating-dominated regions 
(MDR; see Section \ref{mdr}), and also proposed the use of similar species : OH$^+$, OH, H$_2$O$^+$, H$_2$O, and H$_3$O$^+$. 

\subsection{Mechanical Heating}\label{mdr}

\begin{figure}[bh]
  \begin{minipage}[c]{0.6\textwidth}
  \centering{\sf  HC$_{3}$N} 
\includegraphics[scale=0.30, trim= 0 0 0 0, angle = 270]{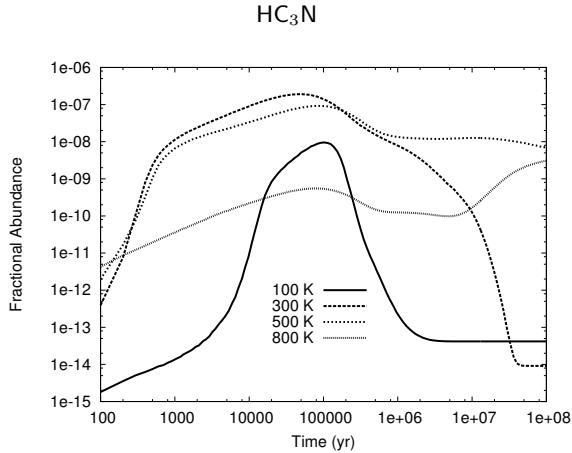}

\end{minipage}
\begin{minipage}[c]{0.4\textwidth}
\vspace{2mm}
\caption{Fractional abundances of HC$_{3}$N at various gas kinetic temperatures calculated with a gas-phase
chemical model including high-temperature reactions in \citet{2010ApJ...721.1570H}.\label{fig:hc3n}}
\end{minipage}
\end{figure}

Increase of temperature can also be caused by ``mechanical heating", shocks or turbulence due to supernova explosions or stellar winds.
 When the kinetic temperature of gas increases, rates of some reactions with high barriers increase significantly ($k \propto exp(-\gamma/T_{kin})$, 
 where $\gamma$ is the barrier of the reaction).
 A well-known example is efficient production of water through accelerated reactions of 
 O + H$_{2}$ $\longrightarrow$OH + H and OH + H$_2$ $\longrightarrow$ H$_2$O + H. 
\citet{2010ApJ...721.1570H} found that this efficient production of water can cause an effective C-rich condition because the dominant form of 
 oxygen become water, and the fractional abundances of O, O$_{2}$ decrease. 
 Reactions between carbon-chains with atomic oxygen are one of the main destruction routes for the destruction of carbon-chain molecules.
 Thus, this lack of atomic oxygen creates an effective C-rich condition, which leads to an enhancement of
  fractional abundances of carbon-containing species such as HCN, HC$_3$N, and C$_2$H$_2$ (Figure \ref{fig:hc3n}). 
\citet{2008A&A...488L...5L} examined the effects of mechanical heating on the line intensities of HCN, HNC, and HCO$^+$ in a steady-state model.
HNC is also known to be sensitive to the elevation of the gas temperature because of a reaction
${\rm HNC + H \longrightarrow HCN + H}$ although the exact value of the barrier is still in debate \citep{2014ApJ...787...74G}.
\citet{2012A&A...542A..65K,2016A&A...595A.124K} extended their work by connecting the mechanical heating rate to the star formation rate, 
and by considering the probability density distribution function.
The work mentioned above assumed constant increase of temperature. However, if the shock heating contribution 
is significant, the temperature fluctuates as a function of time. The case of constant temperature increase and the case of shock
heating may affect reactions with barriers differently, therefore, the chemistry may be different between these two cases.
Such study of temperature fluctuation in the context of GMC needs to be investigated.

Besides the elevation of the temperature, shock waves affect the chemistry
by sputtering ice species from mantles, or by sputtering from grain cores. 
SiO is a well-known example of grain core sputtering caused by strong shocks \citep[e.g., ][]{1992A&A...254..315M}.
For the ice sputtering, CH$_{3}$OH is an example of possible weak shock tracer from ice sputtering 
although non-thermal desorption such as cosmic-ray or UV-induced desorption can also increase methanol in the gas-phase.
At the same time, there is a possibility that methanol gets destroyed by fast shocks ($v_s > 15$ km/s ) \citep{2014MNRAS.440.1844S}.

\section{Astrochemical Observation in Nearby Galaxies}
Statistical studies of molecular lines in various galaxy types can highlight the difference caused by different heating mechanisms.
In fact, some molecular intensities / abundances are suggested to be diagnostics of 
galaxy properties. For example, HCN/HCO$^+$ ratios are proposed to be enhanced in AGN-containing galaxies
as opposed to starburst galaxies \citep{2001ASPC..249..672K,2013PASJ...65..100I}.
In addition, \citep{2011A&A...525A..89A,2015A&A...579A.101A} claimed that CH$_{3}$CCH is enhanced in starburst galaxies than in AGN-containing galaxies,
especially in an old starburst galaxy such as M 82 in their survey of nearby galaxies using IRAM 30-meter telescope. 
Another possible diagnostic molecule is HC$_3$N. \citet{2011A&A...528A..30C} found the enhancement of HC$_3$N/HCN in 
compact obscured galactic nuclei.  
Above mentioned studies done with single-dish observations are ideal to follow-up with high-capability interferometers such as ALMA or NOEMA.
In fact, those high-sensitivity interferometers can discover new molecular tracers. 
Current status of observational efforts is summarized below.

\subsection{GMC-scale Chemistry}\label{sec:gmc}
Although extreme regions of extragalactic sources are interesting, an important benchmark should first be made in the Galaxy for the better interpretation.
Even with ALMA, the beam size to observe nearby galaxies in multiple species are at least a few tens of parsec, about the size of GMCs.
However, previous Galactic astrochemical observations have focused on sub-pc scale star-forming regions.
Thus, before the chemistry in extreme regions are discussed, the chemistry in relatively quiescent regions in a GMC-scale must be understood.
To obtain this ``normal" chemical abundances in a GMC-scale, observational effort has been recently made for Galactic spiral-arm GMCs
such as W49 \citep{2015A&A...577A.127N} in the 1-mm band, Orion B \citep{2017A&A...599A..98P},
 W51 \citep{2017arXiv170700937W}, Orion A \citep{2017arXiv170705352K}, and W3 (OH) (Nishimura et al. submitted) in the 3-mm band.
 One of the important results many of these studies found is that even molecules with high critical densities such as HCN has a rather extended emission.
  Although HCN has been used as a ``dense gas" tracer in many of extragalactic studies \citep[e.g., ][]{2004ApJ...606..271G}, 
 some of the densities of the gas emitting HCN may even be $n \sim 10^3\,$cm$^{-3}$ \citep{2017arXiv170705352K}.
For a template of a non-starburst galactic nucleus, mapping observations of Galactic Center CMZ 
 by \citet{2012MNRAS.419.2961J,2013MNRAS.433..221J}
are also important as benchmarks. Similar large-scale observations in higher frequency bands are needed to 
obtain accurate column densities and to compare the chemistry in
ALMA observations of nearby galaxies.

\subsection{Starburst Galaxies}
Starburst activities in galactic nuclei can be induced by merging of two galaxies, interaction with another galaxy,
and certain galactic structures such as a bar to induce gas inflow into the center of galaxies. 
A connection of the galactic structures and the chemistry is first proposed by \citet{2005ApJ...618..259M} in their multi-species
observations of a starburst galaxy IC 342 in the 3-mm band. They proposed the picture where methanol emission 
tracers shocked regions due to the orbit intersection or due to the density waves, PDR tracers such as CCH is 
enhanced in nuclear starburst regions. They also proposed a similar picture in their ALMA observations of 
a nearby starburst galaxy NGC 253 \citep{2015ApJ...801...63M}. 

\begin{figure*}[bh]
\vspace*{-0.2 cm}
\centering{
  \includegraphics[width=0.70\textwidth, trim = 0 0 0 0]{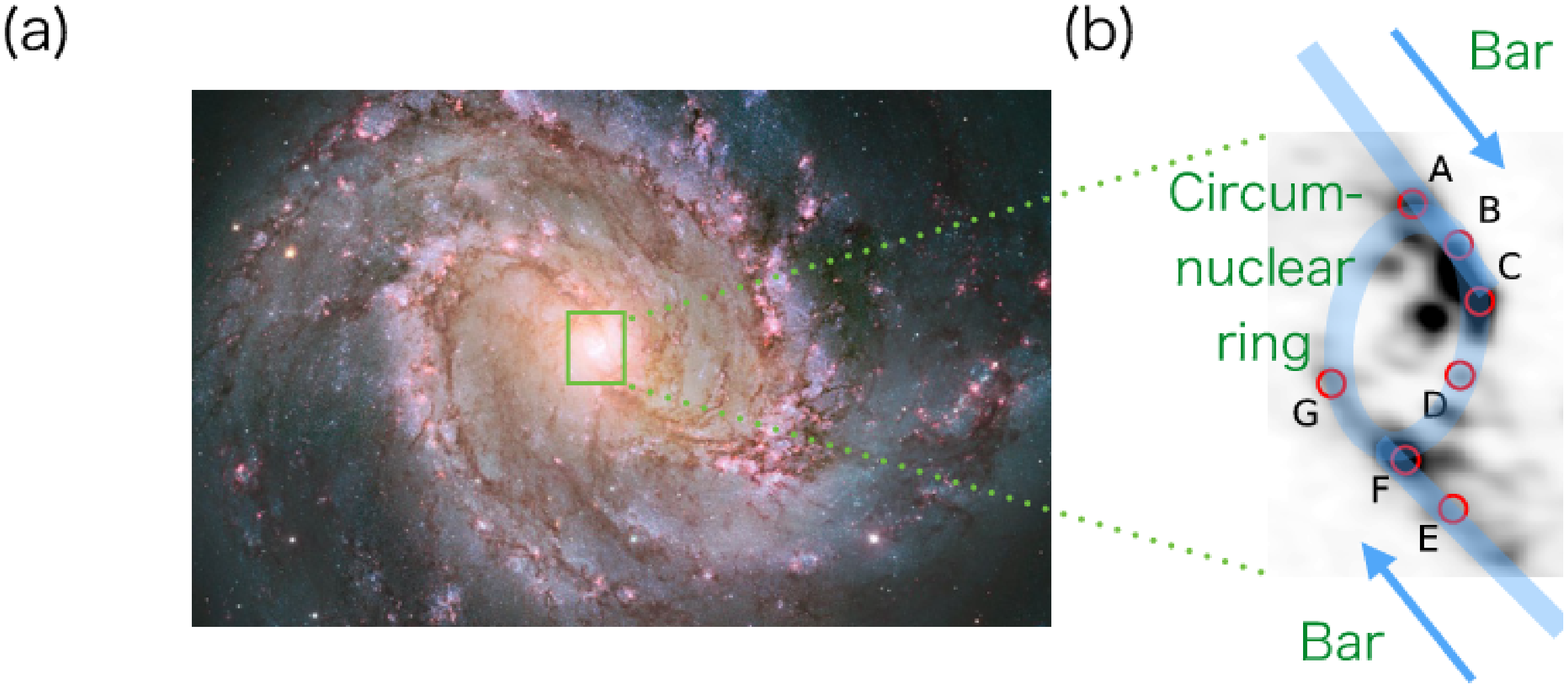} 
 }
\centering{
 
 \includegraphics[width=0.45\textwidth]{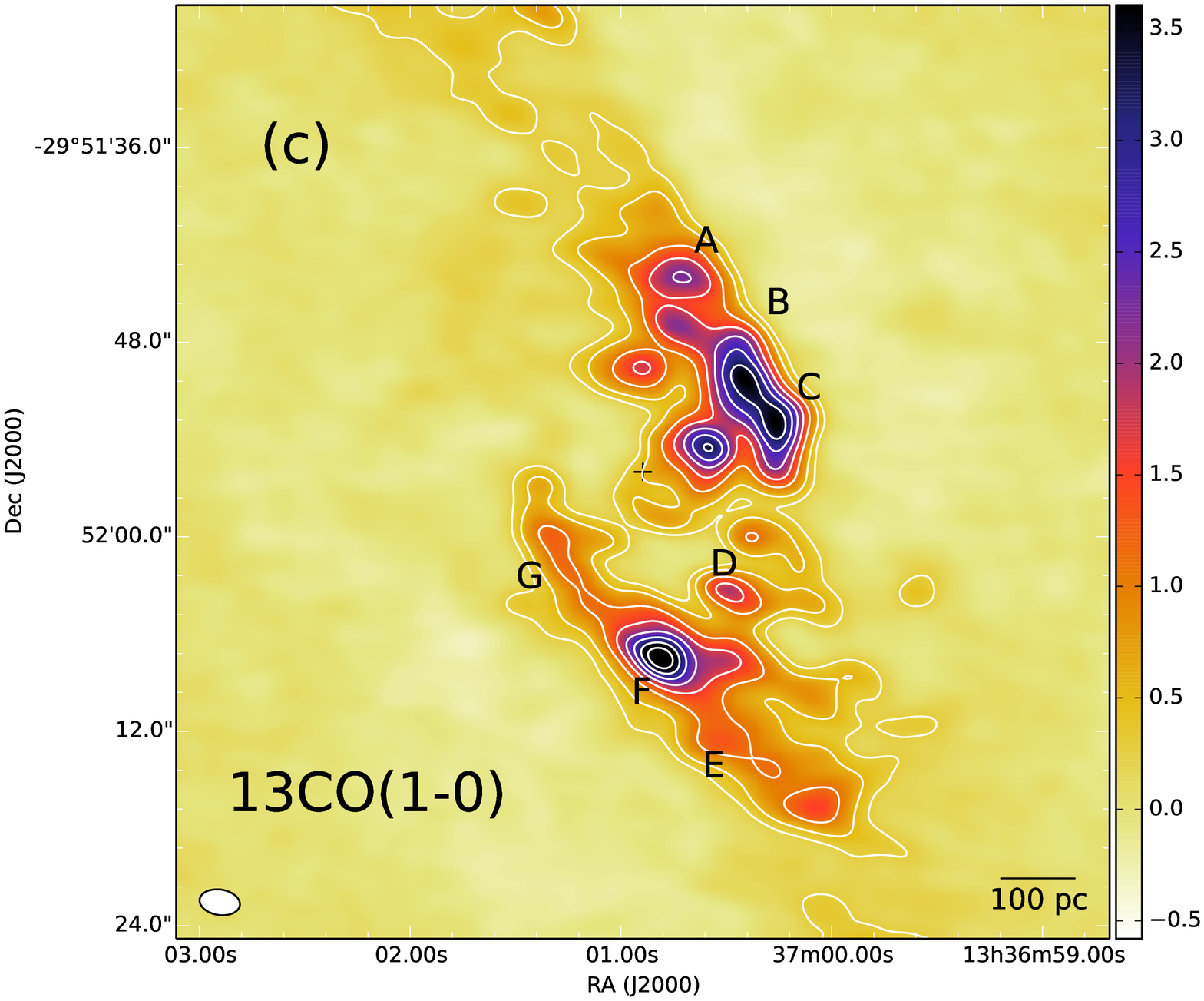} 
\includegraphics[width=0.45\textwidth]{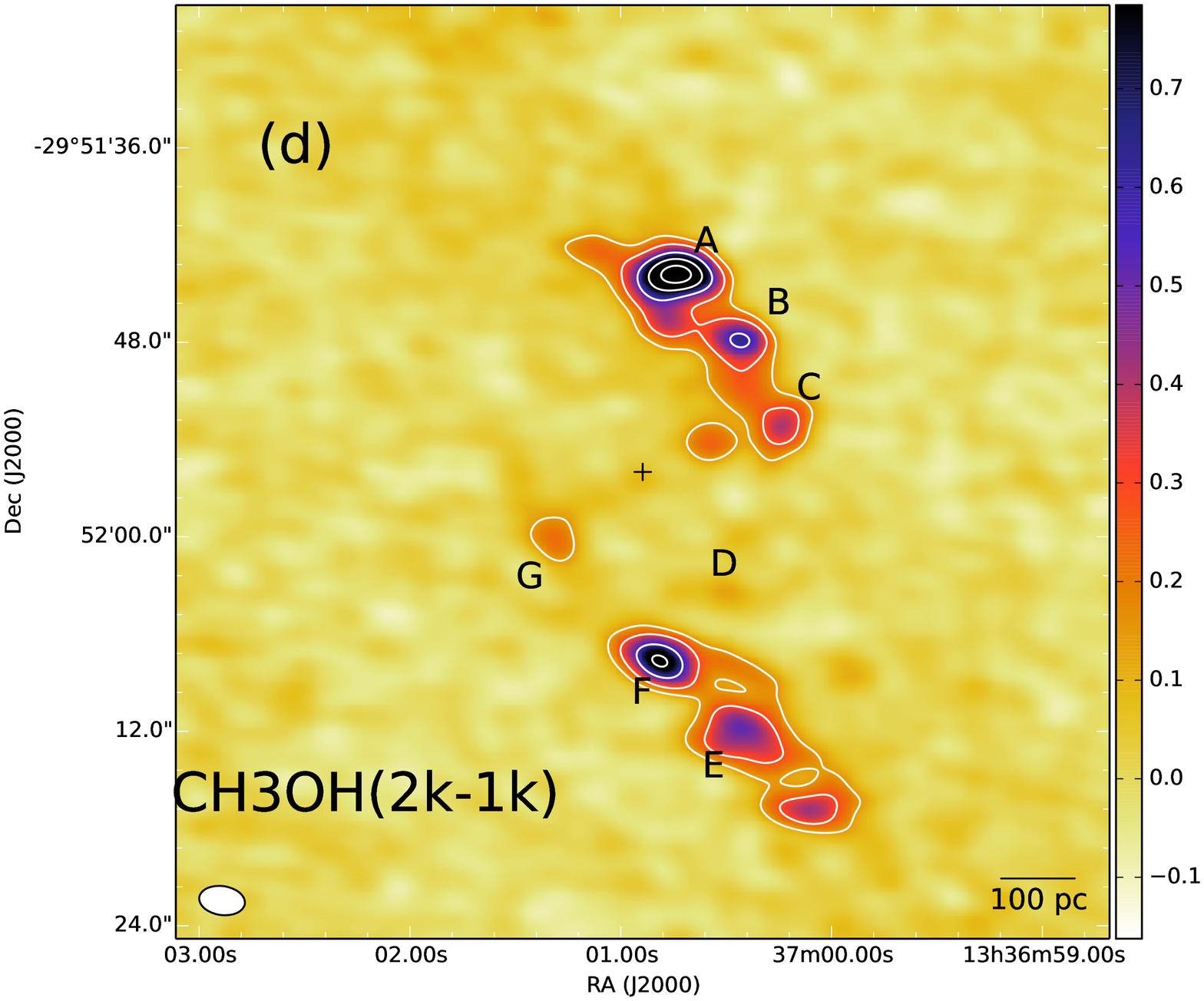} 
 }
 \centering{
\includegraphics[width=0.45\textwidth]{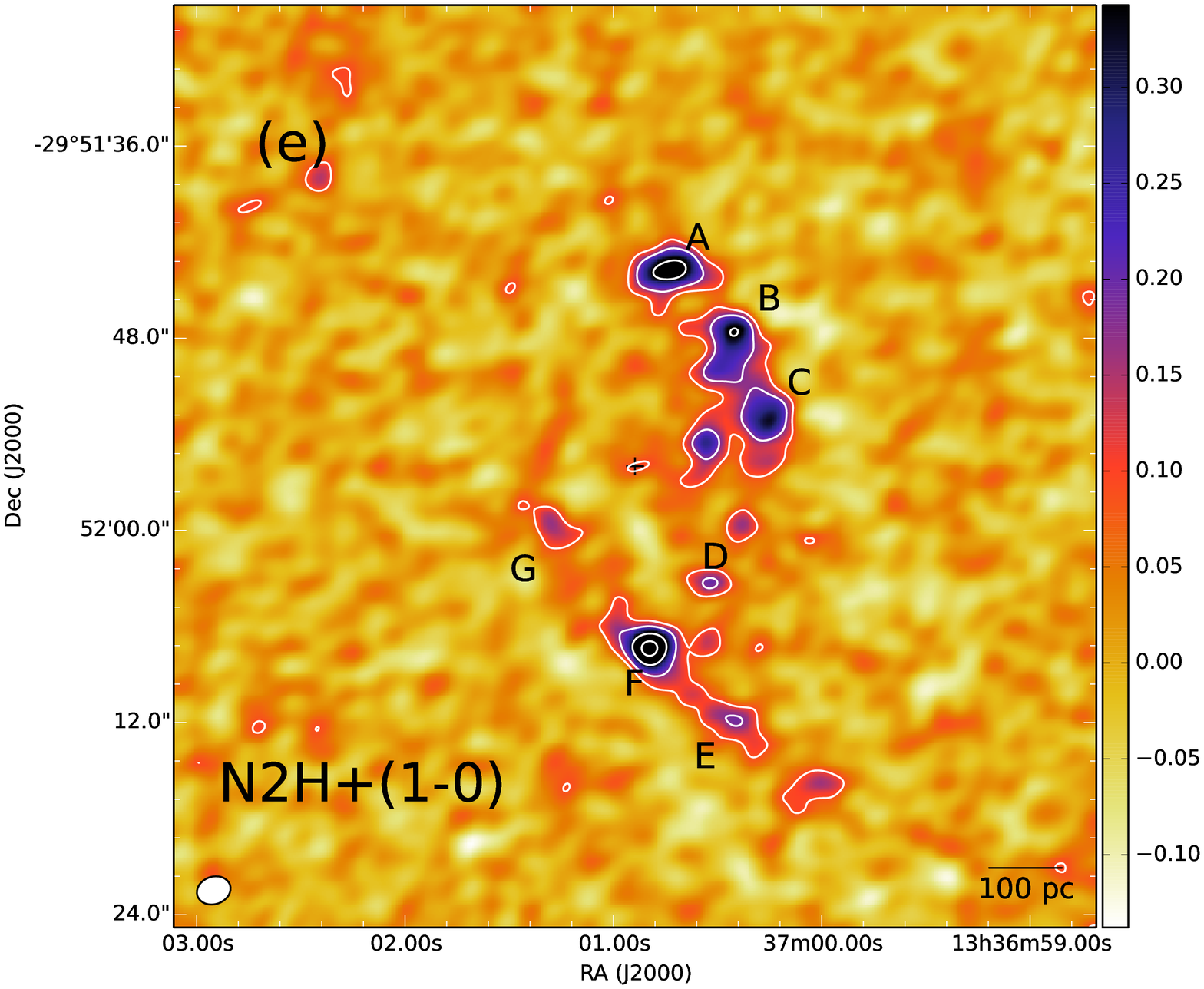} 
 \includegraphics[width=0.45\textwidth]{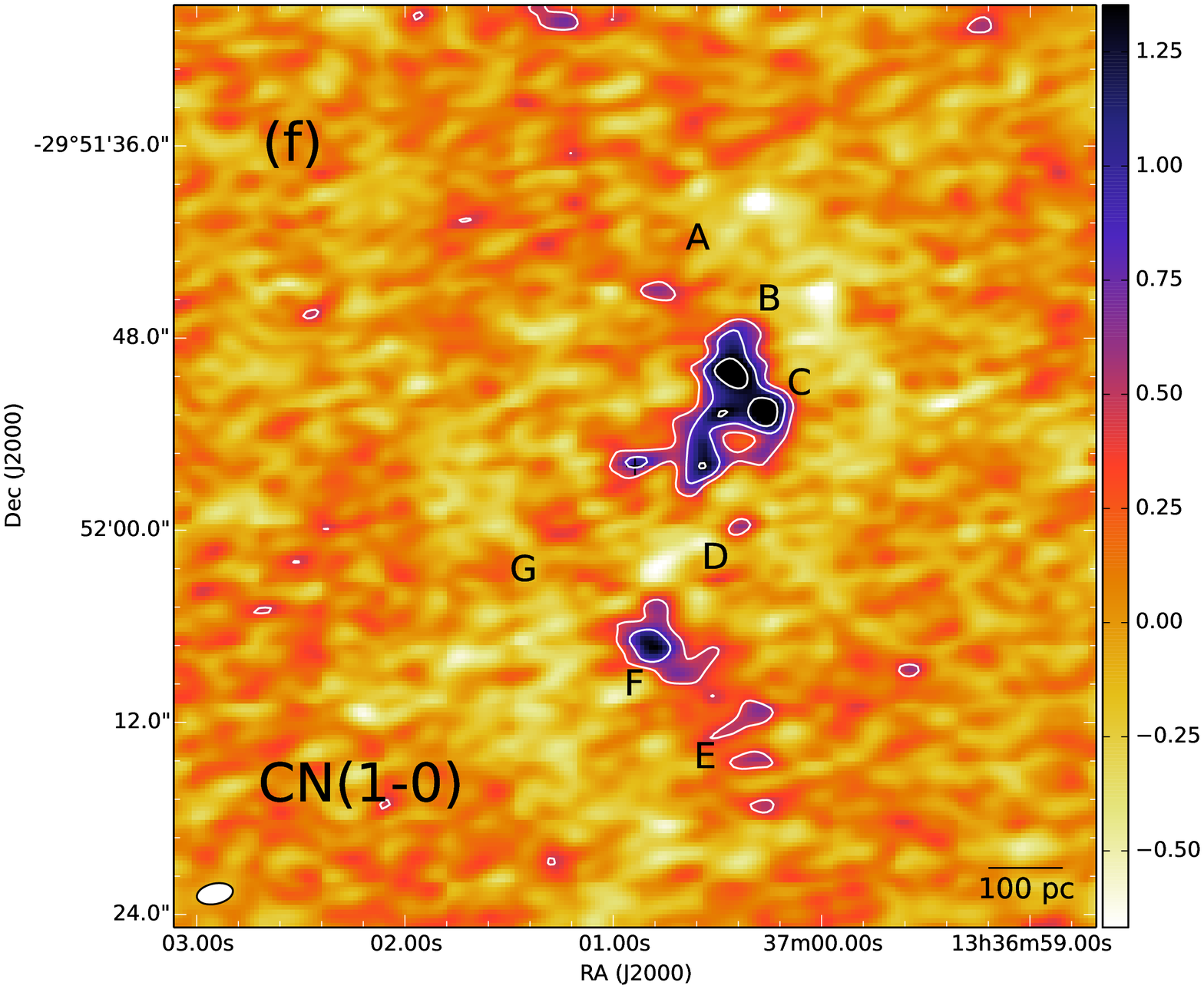} }
 \vspace{1.0 cm}
\caption{(a) $Hubble$ image of M83 (credit: NASA, ESA, and the Hubble Heritage Team (STScI/AURA)) with the area shown in the panel (b) indicated as a rectangle. 
(b) A schematic image of the bar orbit and the orbit of circumnuclear ring overlaid on the $^{13}$CO (1-0) map. Velocity-integrated intensity maps of (c) $^{13}$CO (1-0), (d) CH$_3$OH (2$_k$-1$_k$), (e) N$_2$H$^+$(1-0), and (f) CN($1_{3/2}$-0$_{1/2}$) in the circumnuclear ring of M83
obtained from ALMA data 2015.1.00175.S (PI: Harada) except for CN, which is taken from the ALMA archival data 2011.0.00772.S.
The units of those figures are Jy/beam km/s, and contours are plotted for every $3\sigma$.}
   \label{fig:m83}
\end{figure*}

More recently, Harada et al. (in preparation) observed a nuclear region of a face-on starburst galaxy M83 in multiple molecular tracers.
The observed area includes the regions where the bar orbit intersects with an inner orbit, the circumnuclear ring.
 Those galactic bars are said to promote inflow of the gas into the central regions, and promote star formation \citep{1997ApJ...487..591H}.
In these observations, methanol emission is enhanced at the entering point of gas into the circumnuclear ring (Cloud A in Fig \ref{fig:m83}d)
compared with $^{13}$CO tracing overall molecular gas (Cloud A in Fig \ref{fig:m83}c). 
Although methanol emission can be caused by cosmic-ray induced photo-desorption, it is unlikely that cosmic-ray ionization rate 
is very different in only in two GMCs in the orbit. Methanol emission can also come from hot cores, but such contribution should be 
compact, and methanol emission in the GMC scale is known to be extended \citep[e.g., ][]{2017arXiv170700937W}.
Slightly downstream in the central orbit, N$_2$H$^+$ emission is enhanced (Cloud A \& B in Fig \ref{fig:m83}e).
 N$_2$H$^+$ is normally seen in dense cores in the Galaxy.
Further down in the orbit, there is a peak of CN emission (Cloud B \& C in Fig \ref{fig:m83}f), 
which could be enhanced by PDRs.
From this picture, we propose a scenario where shocks at the orbit intersection compress the gas creating
dense gas clumps eventually to induce star formation. 


\subsection{AGN-containing Galaxies}
An influence of an AGN is initially thought to appear most prominently as X-ray dominated regions of molecular clouds.
Yet, observational trends such as enhanced HCN/HCO$^+$ ratios cannot be explained by XDR models. 
Observational results also seem to indicate that the elevated intensity HCN/HCO$^+$ ratios in AGNs are not from XDRs.
NGC 1068 is a well-studied AGN-containing galaxy.
High-resolution observations of NGC 1068 with ALMA by \citet{2014A&A...567A.125G} and \citet{2014A&A...570A..28V} showed that the intensity
ratio of HCN(1-0)/HCO$^+$(1-0) decreases within the central $\sim 30$ pc.
Also in NGC 1068, observations by \citet{2014PASJ...66...75T} and \citet{2015PASJ...67....8N} detected HC$_3$N, HNCO, 
CN, CS, SO, CH$_3$CN, and CH$_3$OH in their ALMA cycle 0 observations. In their observations, HC$_3$N had a similar fractional abundance to 
that of a Galactic cold dense core TMC-1. 
A relatively large molecule such as HC$_3$N is prone to dissociation by radiation.
This relatively high HC$_3$N fractional abundance means that a few 100-pc scale circumnuclear disk (CND) in NGC 1068 is not a giant XDR,
but contains a significant amount of dense shielded gas from radiation.
A relatively high fractional abundances of CH$_{3}$OH in their observations may result from shocks.
Results by \citet{2010A&A...519A...2G} and \citet{2017A&A...597A..11K} are also consistent in claiming a strong influence of shock.
The elevated intensity HCN/HCO$^+$ ratio in a few 100-pc scale CND is likely to be caused by shocks.
A low-luminosity AGN NGC 1097 also has a high intensity ratio of HCN/HCO$^+$.
In NGC 1097,  \citet{2015A&A...573A.116M} compared the chemical compositions of 
the nuclear region and a kpc-scale circumnuclear ring.
They found elevated abundances of SiO, HNCO, and HC$_{3}$N in the nuclear region, indicating the possibility of shocks.

\subsection{Merging Galaxies}
Merging events can induce star formation by promoting the gas inflow into the central regions.
Such merger events can cause a change in chemical compositions in the interaction regions of 
two galaxies or cause an evolution during the merger stages. 

NGC 4038/4039 (``Antennae" galaxies) are the closest mergers with the distance of 22 Mpc,
whose galactic nuclei are separated by $\sim 7.3$ kpc. Ueda et al. (2016) observed Antennae galaxies
in several molecular species. In their observations, enhancement of CH$_{3}$OH and HNCO was found in a location close to
the northern nucleus. They concluded that this enhancement is caused by galactic scale shocks,
and that other production mechanisms of gas-phase methanol are unlikely.
 VV 114 is a galaxy merger system whose separation of nuclei is 6 kpc, similar to that of Antennae.
 In between the two nuclei of VV 114, there is an overlap region of two galaxies extending about 4 kpc.
 In this overlap region, \citet{2017ApJ...834....6S} found the increased fractional abundance of CH$_{3}$OH
 in a kpc scale. This methanol enhancement was also proposed to be caused by shock 
 due to the interaction of two galaxies.

\begin{figure*}[tp]
\centering{
\includegraphics[scale=0.30, trim= 0 -0 0 0]{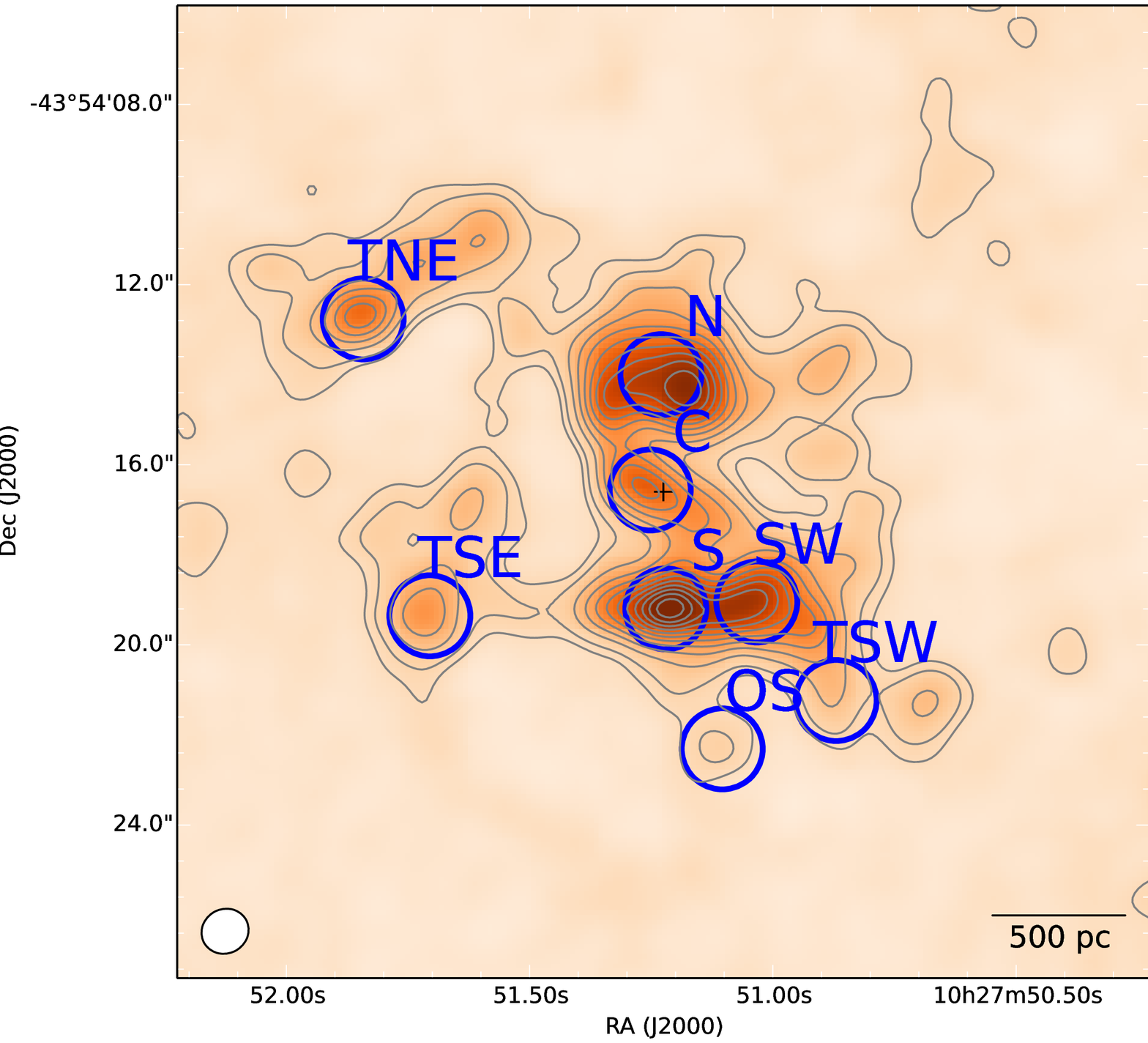}
 \includegraphics[scale=0.30, trim= 0 -100 0 100]{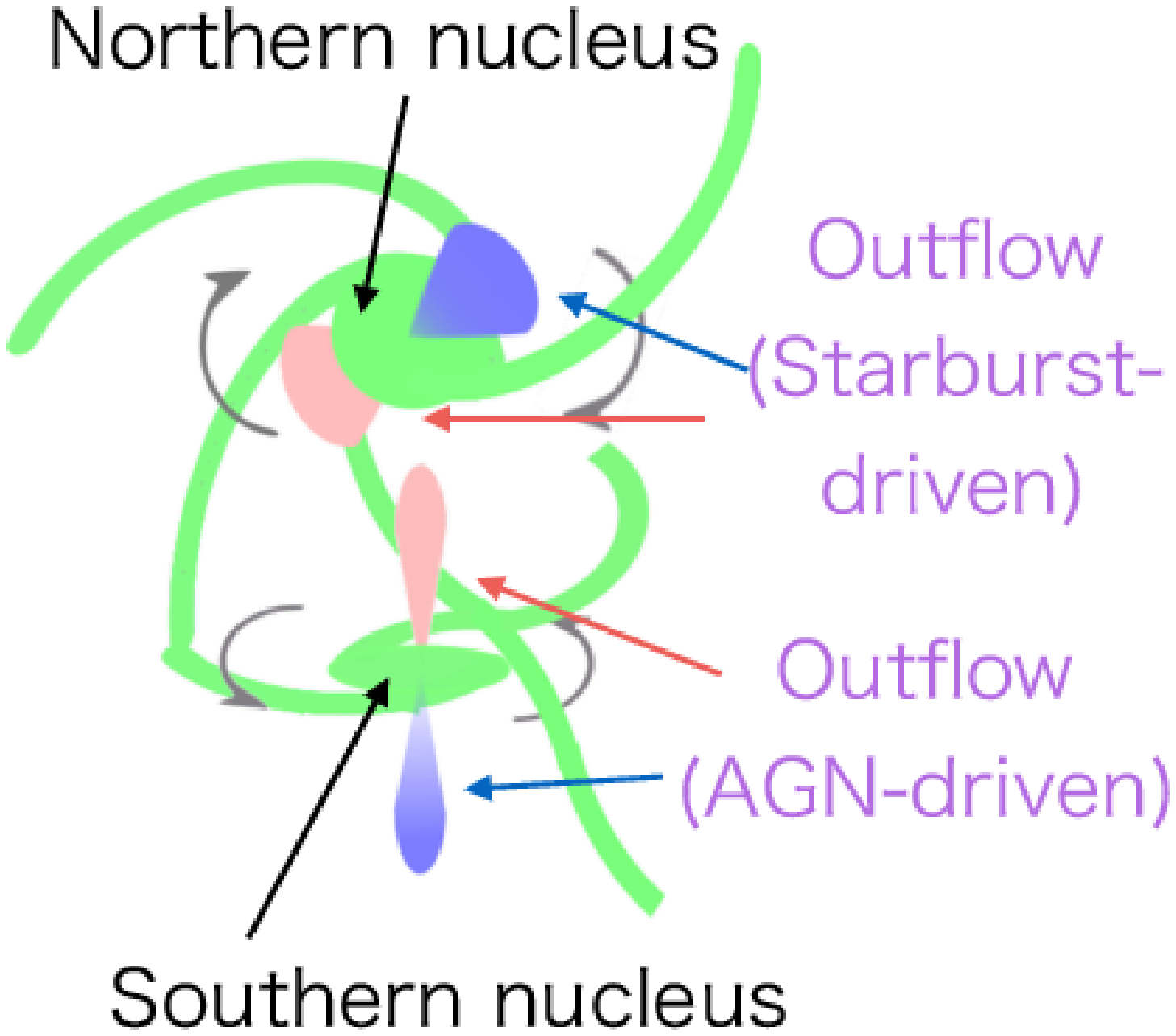}}

\centering{
 \includegraphics[width=0.40\textwidth, trim= 0 0 0 0]{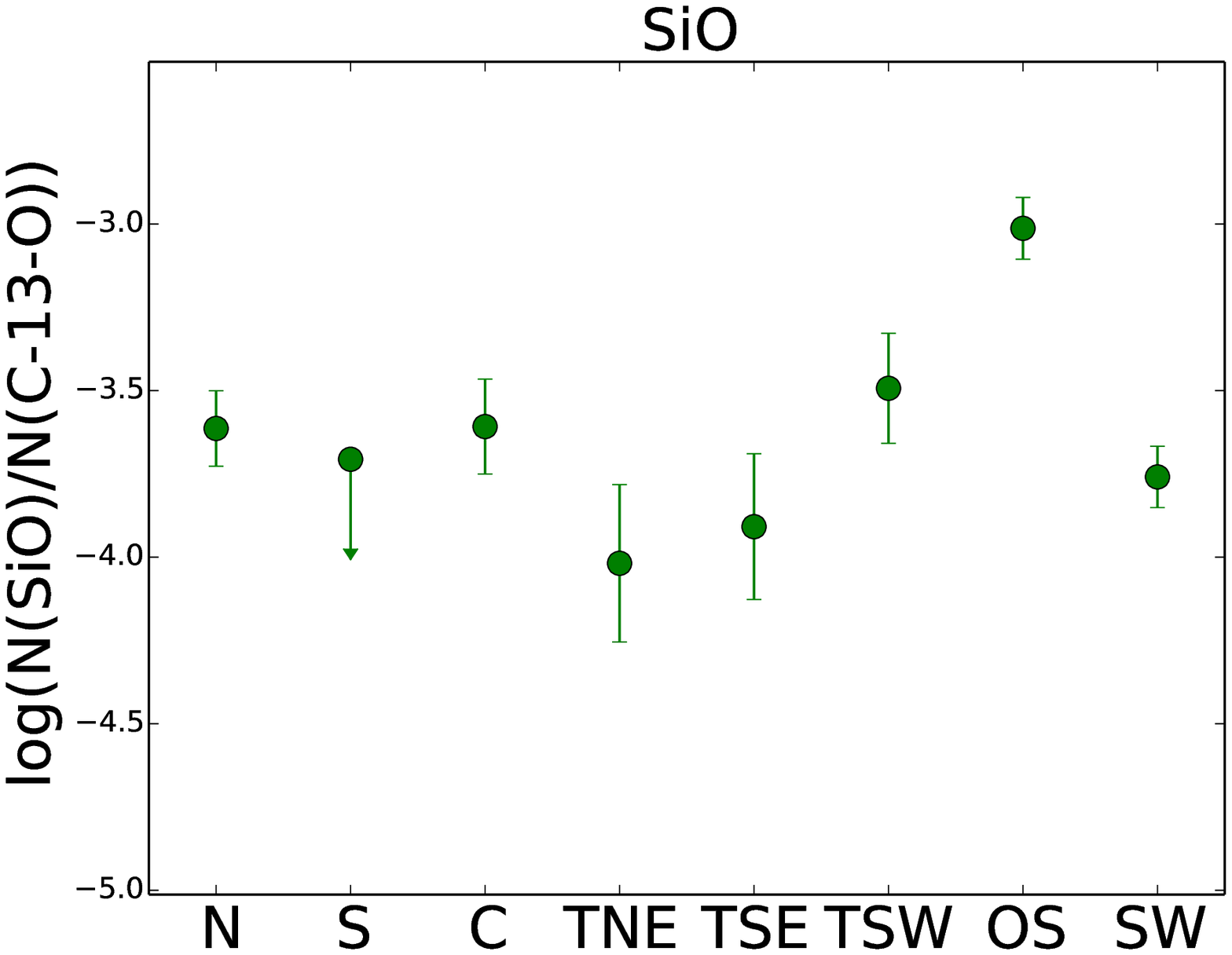}
 \includegraphics[width=0.40\textwidth, trim= 0 0 0 0]{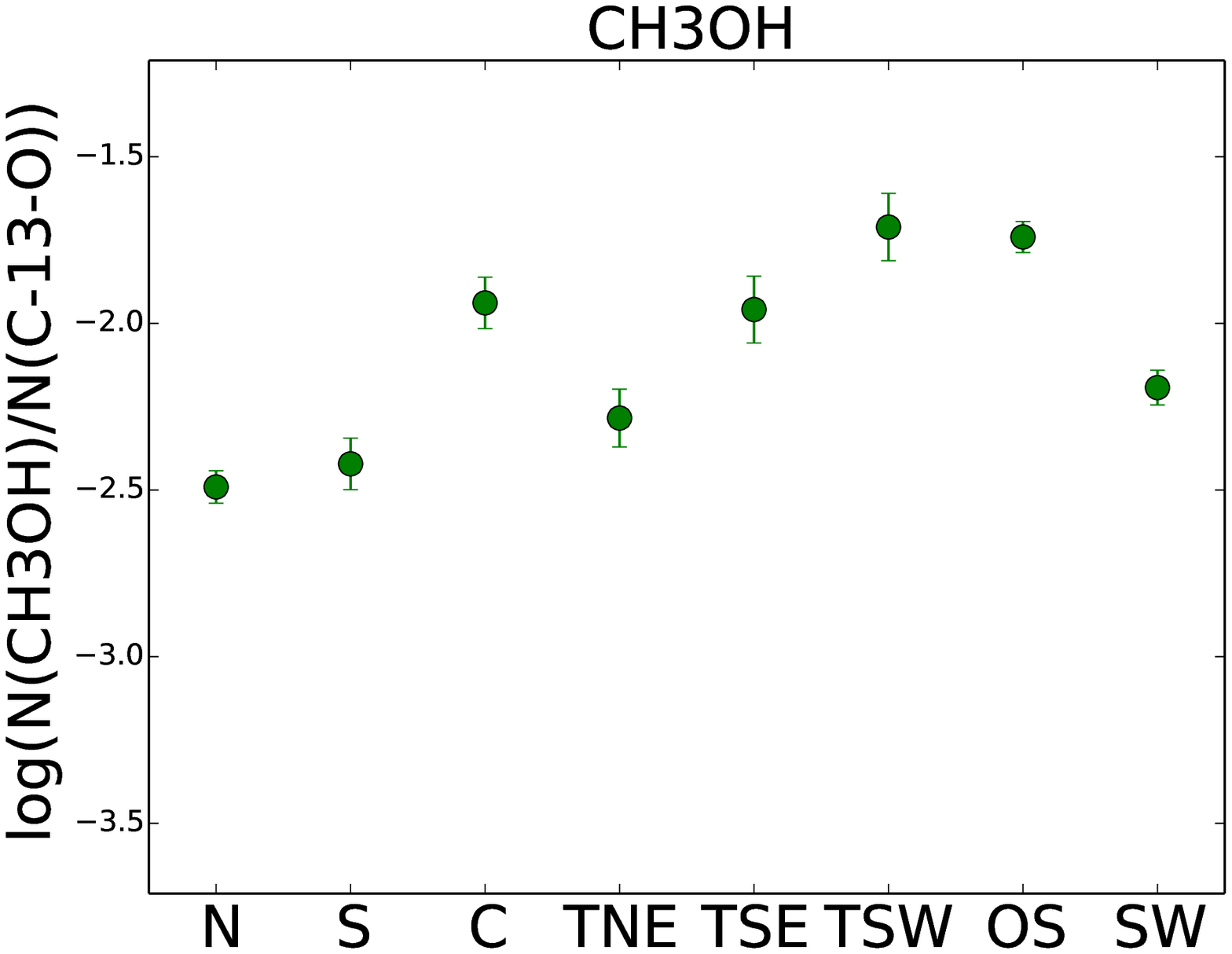}}
\centering{
 \includegraphics[width=0.40\textwidth, trim= 0 -0 0 0]{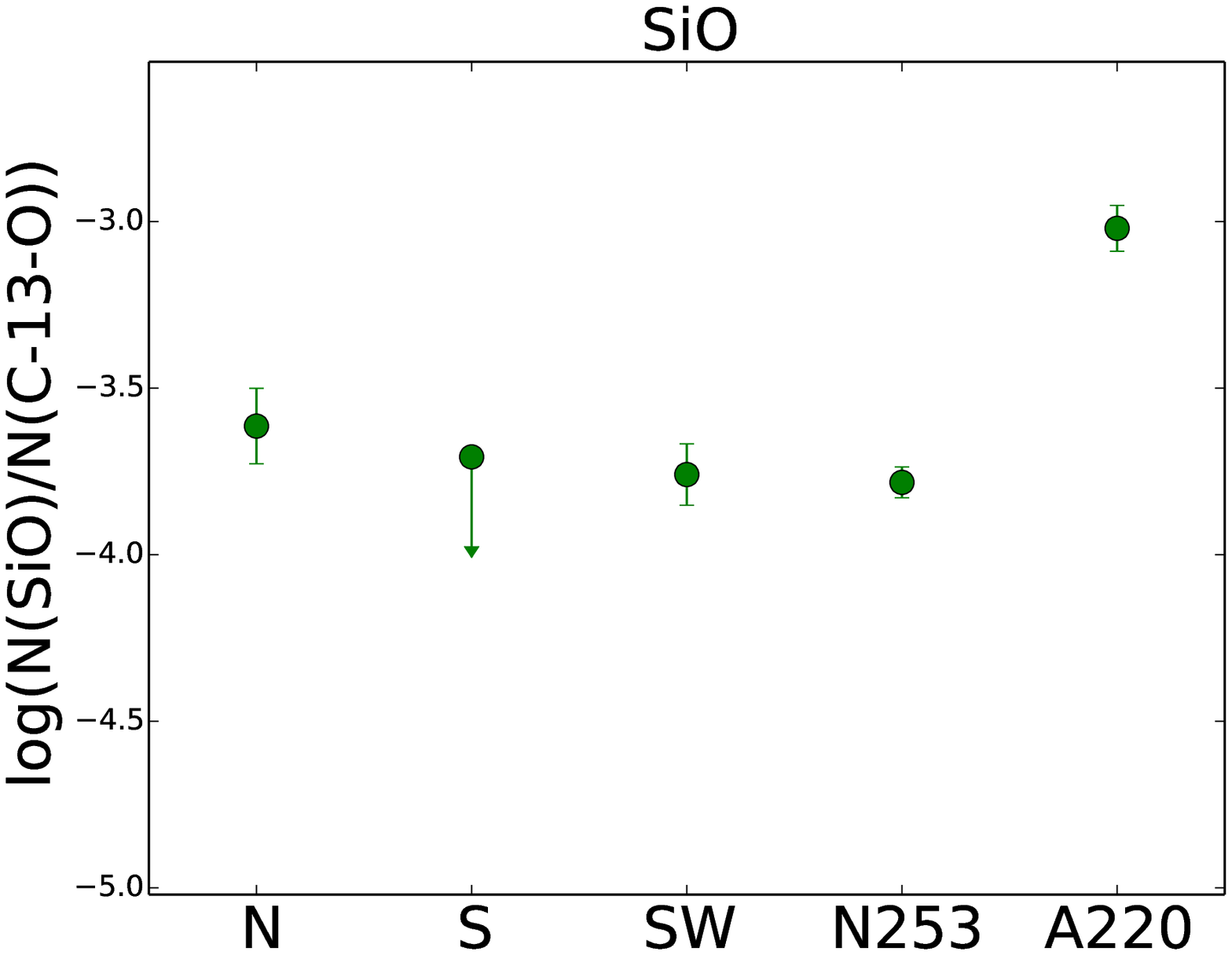}
 \includegraphics[width=0.40\textwidth, trim= 0 0 0 0]{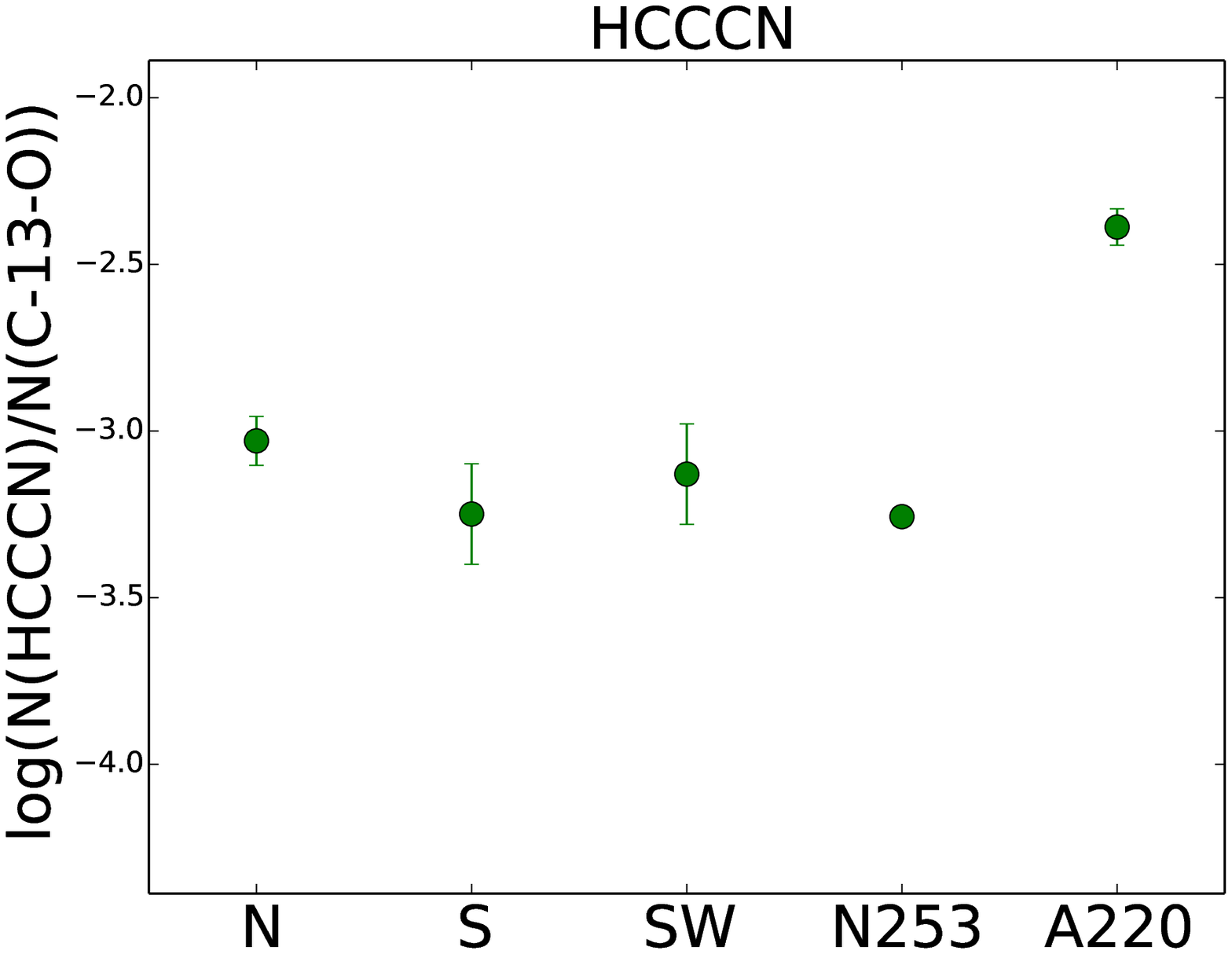}

}
\caption{
{({\it Top left}) A velocity-integrated intensity map of $^{13}$CO(2-1) in NGC 3256 (ALMA data 2015.1.00412.S; PI: Harada).
Each position is labeled as N: Northern nucleus, S: Southern nucleus, C: central position, 
TNE: tidal arm northeast, TSE: tidal arm southeast, TSW: tidal arm southwest, OS: Outflow south, SW: southern nucleus west.
({\it top right}) A cartoon picture of morphology of NGC 3256.
({\it middle left}) Column density ratios of SiO over $^{13}$CO in locations indicated in the top left panel.
({\it middle right}) Same as the middle left panel, but of CH$_{3}$OH.
({\it bottom left}) Column density ratios of SiO over $^{13}$CO in northern nucleus, southern nucleus, southern galaxy west of NGC 3256, a local starburst NGC 253,
and Arp 220. The data of NGC 253 and Arp 220 are taken from \citet{2015A&A...579A.101A}.
({\it bottom right}) Same as the bottom left panel, but of HC$_{3}$N.
}\label{fig:n3256}}
\end{figure*}

NGC 3256 is a late-stage merger with nuclear separation of 850 pc.
Parts of two galaxies are already overlapping, and spiral arms are expected to be influenced by the tidal interactions between the two galaxies (Fig. \ref{fig:n3256} top left).
Because of merger-induced star formation, NGC 3256 has a high infrared luminosity, making it a luminous-infrared galaxy (LIRG).
Another remarkable feature of this merger is molecular outflows from both nuclei.
The molecular outflow from the northern nucleus can be explained only with the energy input from star formation
while the one from the southern nucleus is likely to be caused by an AGN \citep{2014ApJ...797...90S}.
Harada et al. (in preparation) conducted multi-species observations of NGC 3256 with ALMA in the 3-mm and 1.3-mm band.
As we compare the column density ratios over $^{13}$CO or CS in several locations within NGC 3256,
most variations are seen for SiO and CH$_{3}$OH (Fig. \ref{fig:n3256} middle panels). Those species are likely to be tracing the shocked regions.
SiO abundance ratios are particularly enhanced in the outflow regions (OS)
\footnote{In this analysis of the outflow position (OS), we integrated the velocity in the entire velocity range around this galaxy, not only the high-velocity components.
Although it is likely that most of the emission is associated with the outflow, there may be some contamination from components not associated with the outflow.},
and CH$_{3}$OH abundances show similar enhancement. 
Although the desorption mechanisms of CH$_{3}$OH can also be cosmic-ray or UV-photon induced photo-desorption
or chemical desorption, those mechanisms cannot account for increased abundances in particular locations.
Other molecules such as CCH, CN, HC$_3$N, and N$_2$H$^+$ do not have significant variations within NGC 3256.
 When compared with other galactic nuclei, the chemistry of NGC 3256 is very similar to that of NGC 253, 
 despite the fact that NGC 3256 has much higher star formation rates ($15 M_{\odot}$/yr in northern nucleus and $8M_{\odot}$/yr in southern nucleus) than NGC 253
($2 M_{\odot}$/yr). On the other hand, the chemistry of Arp 220 is very different (Fig. \ref{fig:n3256} bottom panels).

Arp 220 is another galaxy merger with even less separation of galactic nuclei ($\sim 400$ pc).
Its chemistry is very distinct. It has high fractional abundances of HC$_3$N
and complex organic molecules compared with local starburst galaxies \citep[][see also Mart\'in et al. in this volume]{2011A&A...527A..36M}.
High excitation condition in this galaxy also allows even detections of vibrationally-excited lines with a lower energy level  $E_{L} > 1000$ K \citep{2016A&A...590A..25M}.

\subsection{Compact Obscured Nuclei}
There are galaxies with very compact, highly obscured, and high-luminosity energy sources in their galactic nuclei.
Because of the obscuration, confirmation of their dominant energy sources, whether they are AGNs or starbursts, 
is difficult.
Arp 220 mentioned above is an example of such galactic nuclei, but
those compact obscured nuclei are not limited to galaxy mergers. 
For example, NGC 4418 does not have an obvious evidence of being a merger, yet its chemistry 
has a great deal of similarity with that of Arp 220.
The fractional abundance of HC$_3$N in NGC 4418 is equivalent to that of Arp 220, 
is more than one order of magnitude compared with local starburst galaxies \citep{2015A&A...582A..91C}.
Strong vibrationally-excited lines are also detected \citep{2010ApJ...725L.228S}.
Although the chemistry in the compact obscured nuclei is unique and interesting, 
understanding the exact mechanism of triggering such chemistry is difficult due to the high degree of obscuration,
and needs further investigation.

\subsection{Outflow Chemistry}
\begin{figure*}[b]
\vspace{-5mm}
\centering{
\includegraphics[width=0.45\textwidth, trim= 0 0 0 0]{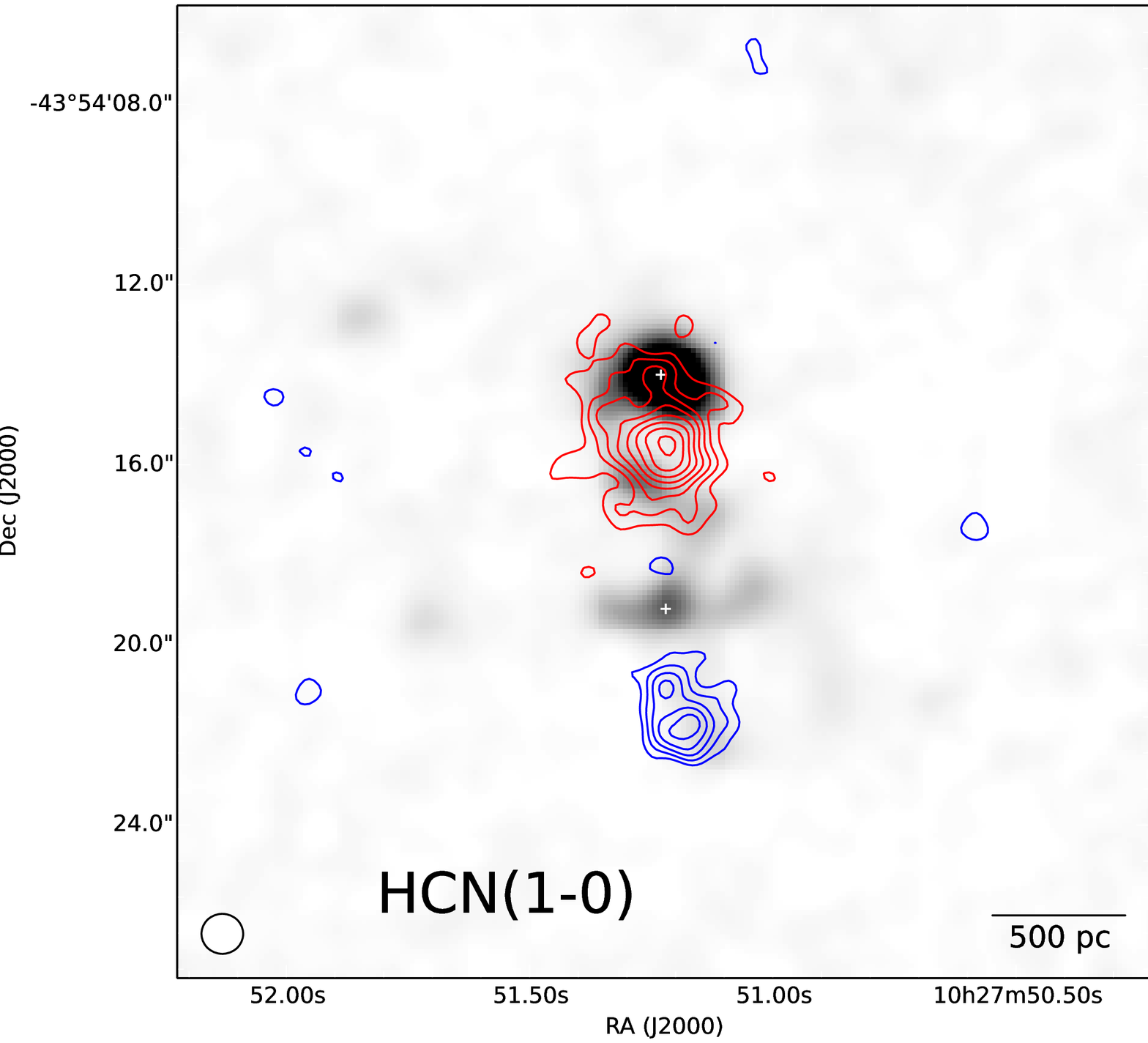}
\includegraphics[width=0.45\textwidth, trim= 0 0 0 0]{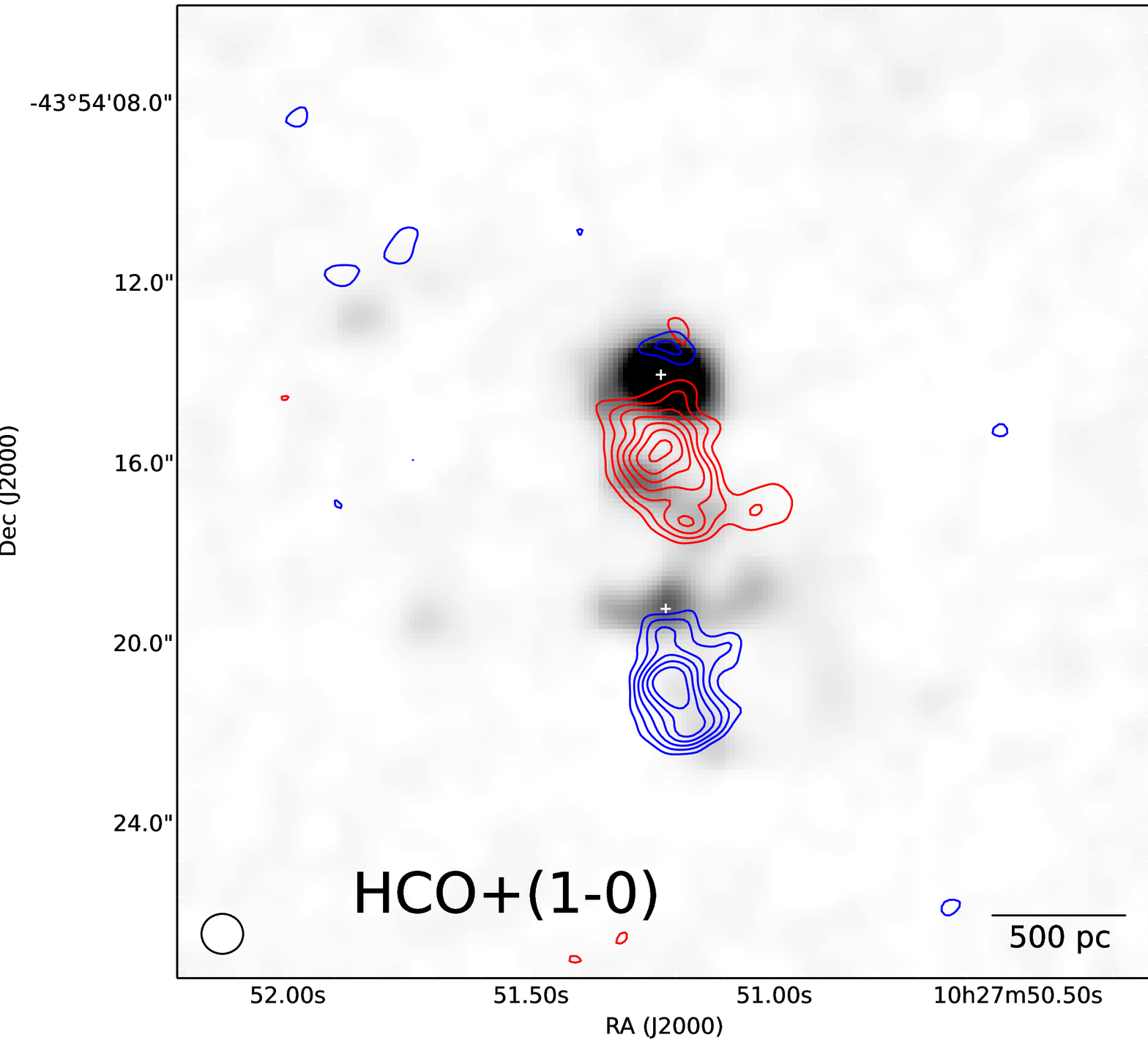}}
\vspace{10mm}
\caption{({\it Left}) An intensity map of  HCN(1-0) integrated for a red-shifted velocity range (+225 to +375 km/s from the systemic velocity) is shown in red contours, 
while a blue-shifted velocity range (-435 to -195 km/s  from the systemic velocity) is shown in blue contours. 
Those velocity ranges are not associated with the nuclear disks of northern and southern galaxies.
({\it Right}) Same as the left panel, but for HCO$^+$(1-0). Data used for these figures are ALMA 2015.1.00412.S,
2016.1.00965.S, and 2015.1.00993.S. Contour levels are starting from $3\sigma$, and plotted for every $1\sigma$.}\label{fig:outflow}
\end{figure*}

Molecular outflow from galactic nuclei can be caused both by starburst and AGN activities,
and can cause quenching of further star formation by expelling the relatively dense and cold gas.
Those outflows are traced by high-velocity components shifted from the systemic velocities of galaxies.
CO outflows are detected in many galaxies so far, both in starburst galaxies and in AGN-containing galaxies.
Although outflow features are usually weak, detections of outflow components are not limited to CO.
In an AGN-containing galaxy NGC 1068, high-velocity components of CN, SiO, HCN, and HCO$^+$ are 
detected \citep{2010A&A...519A...2G,2014A&A...567A.125G}.
Outflow components of HCN and HCO$^+$ lines were detected in Mrk 231, which is thought to be caused by a quasar-driven wind \citep{2012A&A...537A..44A}.
A follow-up work of this Mrk 231 outflow by \citet{2016A&A...587A..15L} found a high abundance ratio of HCN/HCO$^+ \gtrsim 1000$.
This enhancement is likely caused by a shock in the outflow.
In the outflows of NGC 3256 mentioned above, CN \citep{2014ApJ...797...90S}, HCN, and HCO$^+$ (Harada et al. in preparation)
are also detected (Fig. \ref{fig:outflow}).
In a starburst galaxy NGC 253, HCN, CN, HCO$^+$, and CS are also detected in the outflow \citep{2017ApJ...835..265W}.

Those species have higher critical densities than CO, and were unexpected because outflows are thought to be 
generally hot and tenuous.
Detailed study using these molecules would be useful in determining physical and chemical state of outflow.

\section{Summary}
As shown here, ALMA has already provided rich information in astrochemical observations in external galaxies,
and has shown its potential for even further studies. 
However, quantitative interpretation of those observations in connection with starburst or AGN activities are still challenging.
Detailed case studies in a well-studied source or statistical studies of selected molecular lines in various galaxy types 
would observationally help our understanding of extragalactic astrochemistry. 
Although there are many sophisticated chemical models of PDR, XDR, MDR, and CRDR, 
the complexity of GMC-scale observations poses challenge, and needs to be included in future models.


\end{document}